\documentclass[prd,aps,twocolumn,a4paper,floatfix,nofootinbib,preprintnumbers]{revtex4-2}

\usepackage[utf8]{inputenc}
\usepackage{graphicx,psfrag}
\usepackage{mathrsfs}
\usepackage{amsmath,amsfonts,amssymb}
\usepackage{multirow}
\usepackage{diagbox}
\usepackage{comment}
\usepackage{xcolor}
\usepackage{enumerate}
\usepackage{booktabs}
\usepackage{bbm}
\usepackage[normalem]{ulem}

\usepackage{color}
\definecolor{cyan}{rgb}{0,0.9,0.9}
\definecolor{orange}{rgb}{0.9,0.5,0}
\definecolor{magenta}{rgb}{1,0,1}
\definecolor{purple}{rgb}{0.8,0.4,0.8}
\definecolor{gray}{rgb}{0.8242,0.8242,0.8242}
\definecolor{green}{rgb}{0.,0.8,0.}
\definecolor{BerlinU9}{HTML}{F18800}

\usepackage[colorlinks=true]{hyperref}
\hypersetup{
    colorlinks = true,
    linkcolor = BerlinU9,
    urlcolor = BerlinU9,
    linkbordercolor = {white},
    citecolor = BerlinU9,
    }

\begin{document}

\title{Black-hole formation in binary neutron star mergers: The impact of spin on the prompt-collapse scenario}

\author{Federico \surname{Schianchi}$^{1}$}
\author{Maximiliano \surname{Ujevic}$^{2}$}
\author{Anna \surname{Neuweiler}$^{1}$}
\author{Henrique \surname{Gieg}$^{1,2}$}
\author{Ivan \surname{Markin}$^{1}$}
\author{Tim \surname{Dietrich}$^{1,3}$}

\affiliation{${}^1$ Institute for Physics and Astronomy, University of Potsdam, Haus 28, Karl-Liebknecht-Str. 24/25, 14476, Potsdam, Germany}
\affiliation{${}^2$ Centro de Ci\^encias Naturais e Humanas, Universidade Federal do ABC, 09210-170, Santo Andr\'e, S\~ao Paulo, Brazil}
\affiliation{${}^3$ Max Planck Institute for Gravitational Physics (Albert Einstein Institute), Am M\"uhlenberg 1, Potsdam 14476, Germany
}

\date{\today}

\begin{abstract}

Accurate modeling of the multi-messenger signatures connected to binary neutron star mergers requires proper knowledge on the final remnant's fate and the conditions under which black holes (BHs) can form in such mergers. 
In this article, we use a suite of 84 numerical-relativity simulations in 28 different physical setups to explore the impact of the individual stars' spin on the merger outcome and on the early postmerger dynamics. We find that for setups close to the prompt-collapse threshold, the stars' intrinsic spin significantly changes the lifespan of the remnant before collapse and that the mass of the debris disk surrounding the BH is also altered. To enable a better understanding of BH formation, we check if there is at least a theoretical chance of observing densities that are above the maximum density allowed in a stable isolated neutron star, and we investigate the importance of different pressure contributions on the evolution of the postmerger remnant and BH formation. 

\end{abstract}

\maketitle

\section{Introduction}
\label{sec:introduction}

Neutron stars (NSs) are among the most compact objects known in our Universe and they allow us to probe matter under the most extreme conditions close to the edge of black hole (BH) formation, see e.g., Refs.~\cite{Lattimer:2000nx,Lattimer:2004pg}. In this regard, precise knowledge about the onset of gravitational collapse and the maximum mass of NSs can provide important constraints on the properties of supranuclear-dense matter. Typically, such constraints on the maximum mass of NSs are derived assuming that the considered stars are non-rotating and cold. In this case, the Tolmann-Oppenheimer-Volkoff (TOV) equations~\cite{Tolman:1939jz,Oppenheimer:1939ne} allow us to compute the maximum mass $M_{\rm TOV}$ once a microphysical equation of state (EOS) is known. On the contrary, $M_{\rm TOV}$ can also be used to determine the microphysical EOS. Hence, nuclear physics interactions and properties can be tested and predicted with the help of astrophysical observations. In general, if the star is rotating, the maximum supported mass increases by about 25\% if the star is rigidly rotating~\cite{1987ApJ...314..594F}, i.e., if the angular frequency across the star is constant, and if the star is rotating at the Kepler limit. Even more massive stars might be supported through differential rotation~\cite{Baumgarte:1999cq}. \par

The maximum mass of an NS can be constrained through different astrophysical observations. On the one hand, the observation of NSs either through X-ray (e.g., PSR J0952-0607~\cite{Romani:2022jhd}) or radio measurements (e.g., PSR J1614-2230~\cite{Demorest:2010bx,NANOGrav:2017wvv}, PSR J0740+6620~\cite{NANOGrav:2019jur,Fonseca:2021wxt}) provide a lower bound on the maximum mass of NSs. On the other hand, the observed onset of BH formation either through accretion or during a binary neutron star (BNS) merger provides an alternative way of constraining $M_{\rm TOV}$. Following the detection of GW170817~\cite{LIGOScientific:2017vwq,LIGOScientific:2017ync}, there has been an increasing number of studies attempting to determine the maximum NS mass based on the assumption that GW170817's remnant collapsed into a BH \cite{vanPutten:2022xyx}. This assumption is supported by the observation of both the gamma-ray burst and kilonova emission. Based on these assumptions, several groups have derived upper bounds on the maximum mass of NS, e.g.,~\cite{Margalit:2017dij,Rezzolla:2017aly,Ruiz:2017due,Shibata:2019ctb,Dietrich:2020efo,Fan:2023spm}. \par

One possibility to set the upper bound on the maximum mass of NSs is to determine the prompt-collapse threshold mass, i.e., the mass under which a BNS merger leads to an immediate formation of a BH after the merger ($\lesssim 2\rm ms$)~\cite{Bauswein:2013jpa}. Over the years, there have been several proposals on how the threshold mass is connected to $M_{\rm TOV}$, e.g., \cite{Bauswein:2013jpa,Kolsch:2021lub,Tootle:2021umi,Kashyap:2021wzs,Tootle:2021umi,Ecker:2024kzs}. To our knowledge, most of the works (except for Tootle et al.~\cite{Tootle:2021umi}) focused on irrotational configurations. Hence, the influence of the stars' intrinsic rotation on the prompt BH formation has not been investigated in detail. More generally, numerical-relativity simulations of binary neutron stars with intrinsic spins are still relatively limited, e.g., \cite{Bernuzzi:2013rza,Kastaun:2014fna,Dietrich:2016lyp,Dietrich:2017xqb,Dietrich:2018upm,Dietrich:2018upm,Most:2019pac,Dudi:2021wcf,East:2019lbk,10.1093/mnras/stae454}. To pursue this further investigation, we performed a total of $84$ numerical-relativity simulations with various resolutions, studying $28$ different physical setups ($22$ spinning and $6$ non-spinning setups). \par

The article is structured as follows: In Sec.~\ref{sec:setups}, we briefly describe the setups that we simulate and the methods that we use. In Sec.~\ref{sec:results}, we discuss the results of our simulations, starting with a quantitative discussion of the merger dynamics, an investigation of the apparent-horizon formation, and a discussion of the pressure evolution in the postmerger remnant. We then continue by computing the approximate collapse time and presenting the emitted gravitational-wave (GW) energy and luminosity, as well as the disk masses. We conclude in Sec.~\ref{sec:conclusion}. \par

Throughout the article, we use geometric units and set $M_\odot=c=G=1$, unless otherwise stated. 

\section{Setup} 
\label{sec:setups}

\subsection{Methods}
\label{sec:method}

\paragraph*{Initial Data:} 
The initial configurations simulated in this work are computed with the pseudo-spectral code SGRID~\cite{Tichy:2009yr,Tichy:2012rp,Dietrich:2015pxa,Tichy:2019ouu}. SGRID employs surface fitting coordinates to solve the Einstein Constraint Equations using the extended conformal thin sandwich formulation~\cite{York:1998hy,Pfeiffer:2002iy}. For the construction of spinning BNSs, we use the constant rotational velocity approach introduced by Tichy in Refs.~\cite{Tichy:2011gw,Tichy:2012rp}. 

\paragraph*{Dynamical Evolution:} 
The BAM code~\cite{Bruegmann:2006ulg,Thierfelder:2011yi,Dietrich:2015iva,Bernuzzi:2016pie,Dietrich:2018phi} is used for the dynamical evolution of matter and spacetime fields. For the latter, we employ the Z4c formulation of the field equations of General Relativity (GR) with constraint damping terms~\cite{Bernuzzi:2009ex,Hilditch:2012fp} along with the moving punctures gauge (1+log-slicing and gamma-driver shift conditions~\cite{Bona:1994a,Alcubierre:2002kk,vanMeter:2006vi}). The matter variables are evolved using the Valencia formulation of general-relativistic hydrodynamics (GRHD)~\cite{Marti:1991wi,Banyuls:1997zz,Anton:2005gi}. In this article, we restrict ourselves to pure GRHD, without including neutrino interactions and magnetic fields. Furthermore, we use zero-temperature piecewise-polytropic equations of state (EOSs) following~\cite{Read:2008iy} and incorporate thermal effects following the prescription of \cite{Bauswein:2010dn}, i.e., adding a thermal contribution to the total pressure in the form of $P_{\rm th} = (\Gamma_{\rm th} - 1) \rho \epsilon_{\rm th}$ with $\Gamma_{\rm th} = 1.75$, $\rho$ being the rest-mass energy density, and $\epsilon_{\mathrm{th}}$ is the thermal contribution to the specific internal energy. \par

BAM's grid consists of cell-centered nested grids with $L$ refinement levels labeled by $l=0,...,L-1$. Each level $l$ contains one or more Cartesian boxes with a constant grid spacing $h_l$ and $n$ (or $n_{\rm mv}$ for the inner, moving boxes) points per direction. Due to the 2:1-refinement strategy, the resolution in each level is given as $h_l = h_0/2^l$. Inner levels with $l \geq l_{\rm mv}$ move dynamically, following the motion of the NSs or BHs. For this work, we employ $L=7$ and $l_{\rm mv}=2$. Each physical setup was simulated with three different grid resolutions R1, R2, and R3, corresponding to 96, 128, and 160 grid points per direction on the finest level, respectively. The finest level comprises two boxes, each one entirely covering each star. For R3, this corresponds to a grid spacing $h_{L-1}$ on the finest level of $134\,$m for SLy and $170\,$m for H4.\par

For the time evolution, we use the method of lines with a fourth-order Runge-Kutta scheme and a Berger-Oliger algorithm for the refinement levels. The spacetime sector uses a finite difference scheme with centered fourth-order stencils to compute spatial derivatives. Hydrodynamical variables are instead modeled by a finite volumes formalism with high-resolution shock-capturing schemes to compute numerical fluxes between cells. In this case, we use the WENOZ \cite{Borges:2008} reconstruction of characteristic fields together with a componentwise local Lax Friedrich Riemann solver~\cite{Nessyahu:1990, Kurganov:2000}, following the implementation of \cite{Bernuzzi:2016pie}. Finally, the conservative adaptive mesh refinement strategy implemented in~\cite{Dietrich:2015iva} is employed to guarantee the conservation of baryonic mass, energy and momentum.

\subsection{Configurations}
\label{sec:configurations}

In this work, we consider two different piecewise-polytropic EOSs to describe the NSs, SLy~\cite{Douchin:2001sv}  and H4~\cite{Lackey:2005tk}. These EOSs support single stars in isolation with maximum gravitational masses of $2.06M_{\odot}$ and $2.03M_{\odot}$, respectively. For each EOS, we select three different mass-ratio combinations of the binary $q=M^{\rm A}/M^{\rm B}$, $q = \{1.000, 1.375, 1.625\}$, each with different total gravitational mass $M=M^{\rm A} + M^{\rm B}$, where $M^{\rm A,B}$ are the gravitational masses of the isolated stars. Total masses have been chosen in a way to be the closest to the prompt-collapse mass-threshold of the ones tested in \cite{Kolsch:2021lub} for a given EOS and mass-ratio. In each setup, the individual NS spins, $\chi^{\rm A}$ and $\chi^{\rm B}$, are aligned, antialigned or null with respect to the orbital angular momentum. In particular, we considered $6$ irrotational setups, $12$ setups with orbit-aligned spins for both stars and $6$ setups with orbit-antialigned spins for both stars. Additionally, for the highest mass ratio of each EOS, two more spin-antialigned setups were constructed to have zero effective spin, i.e., $\tilde{\chi} = (M^{\rm A} \chi^{\rm A} + M^{\rm B} \chi^{\rm B})/(M^{\rm A}+M^{\rm B})=0$. However, most of the comparisons with literature results are carried out using the total spin $\overline{\chi} = \chi^{\rm A} + \chi^{\rm B}$.\par 

In total, $28$ different configurations are considered in this work. 

\section{Results} 
\label{sec:results}

To investigate spin effects on the BH formation in BNS simulations, we begin our discussion with a description of the BH formation and the maximum densities that are probeable within BNS mergers (Sec.~\ref{subsec:max_density}). We continue with a discussion about individual pressure components that influence the stability of the remnant (Sec.~\ref{subsec:pressure}), the spin influence on the collapse time, the GW energy and luminosity, and the disk mass. In addition, we summarize the main results of our simulations in Tables~\ref{tab:H4_aligned}, \ref{tab:SLy_aligned}, \ref{tab:anti-aligned} given in Appendix~\ref{app:tables}.

\subsection{Black Hole Formation and Maximum Density}
\label{subsec:max_density}

\begin{figure*}[htp]
    \centering
    \includegraphics[width=\linewidth]{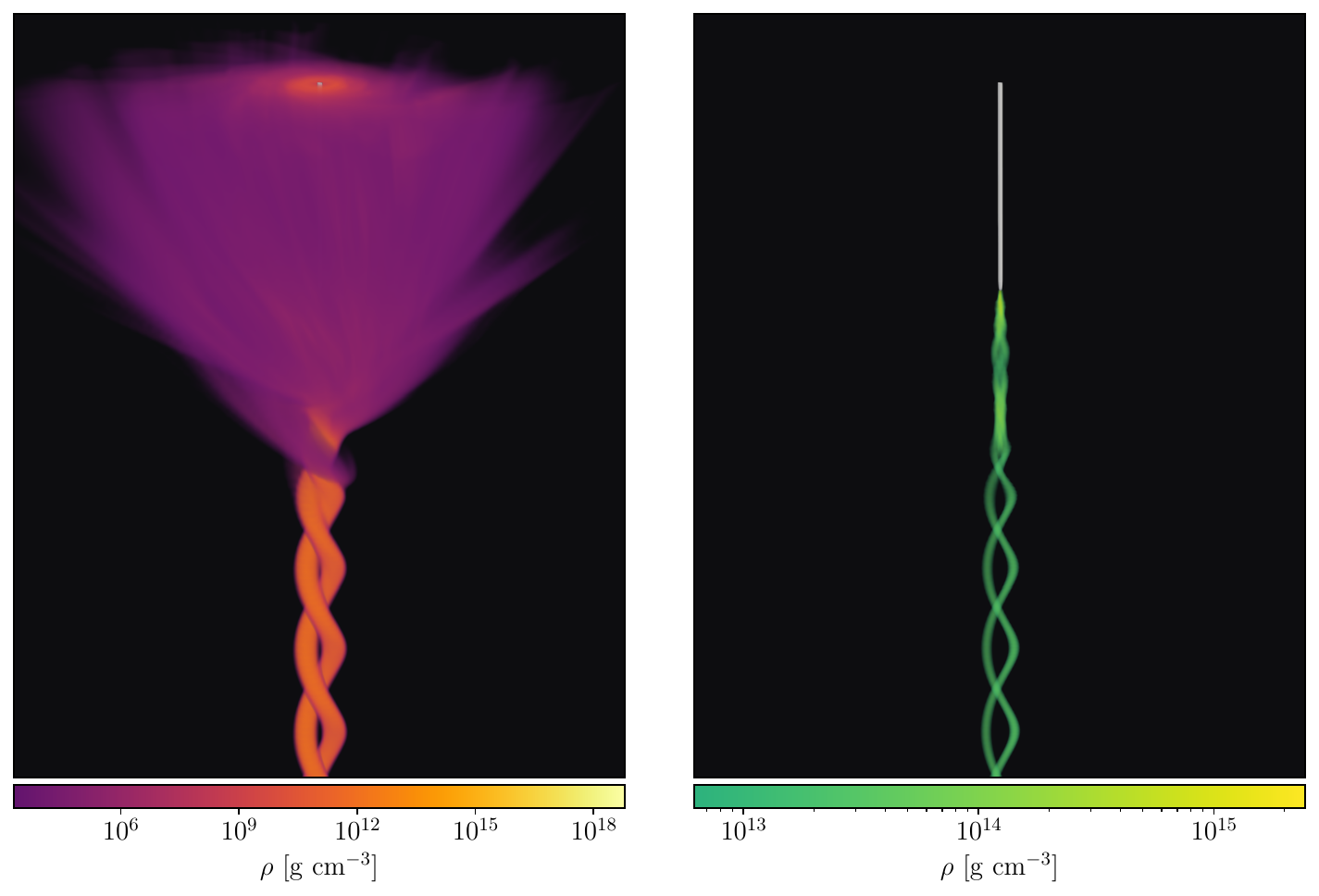}
    \caption{Pants diagram -- vertically stacked slices of matter-density in $x$-$y$ plane with the simulation time progressing upwards until $4.83$~ms after the merger and capturing the last two orbits before the merger. The apparent horizon is represented as a gray surface. The data is for the equal mass simulation with H4 EOS and $\chi^A=\chi^B=-0.1$ at resolution R3.
    \textit{Left panel}: The color map captures a wide range of density, making the ejecta visible. \textit{Right panel}: the color coding captures only a range of high densities, making the evolution of the NS cores and the formation of apparent horizon visible.}
    \label{fig:pants}
\end{figure*}

\textit{Qualitative Discussion:}
To enable a qualitative discussion about the binary dynamics, we present a `pants' visualization similar to the representation commonly used for binary BHs, e.g.,~\cite{Matzner:1995ib}. We plot two-dimensional slices of the density in the orbital plane, stacked vertically with time progressing upwards. Such a diagram provides an overview of the time evolution of the system. In the left panel of Fig.~\ref{fig:pants}, the two NSs, represented by the two tubes, orbit around each other, and at the merger, a fraction of the material becomes unbound (depicted in purple). At the end of the simulation, a BH is found at the center of the remnant. In the right panel of Fig.~\ref{fig:pants}, another color map is used to focus on the evolution of the NS cores ($\rho > 6.17 \times 10^{12}\,\mathrm{g\,cm^{-3}}$) and the subsequent formation of a BH (gray surface). For a recent in-depth, gauge-independent study about black hole formation and prompt-collapse scenarios, we refer to the work of Ecker et al.~\cite{Ecker:2024kzs}. \par

The dynamics and formation of the BH depend on the system parameters such as the masses, spins, and the EOS of the NSs. In this article, we focus on the BH formation, but we refer to, e.g., Refs.~\cite{Hotokezaka:2013iia,Dietrich:2016hky,Dietrich:2016lyp,Kastaun:2016elu,Kawamura:2016nmk,Hanauske:2016gia,Takami:2014tva} for a selection of numerical-relativity studies that investigated the influence of different system parameters on the binary dynamics. \\

\begin{figure*}[htpb]
\centering
\includegraphics[width=\linewidth]{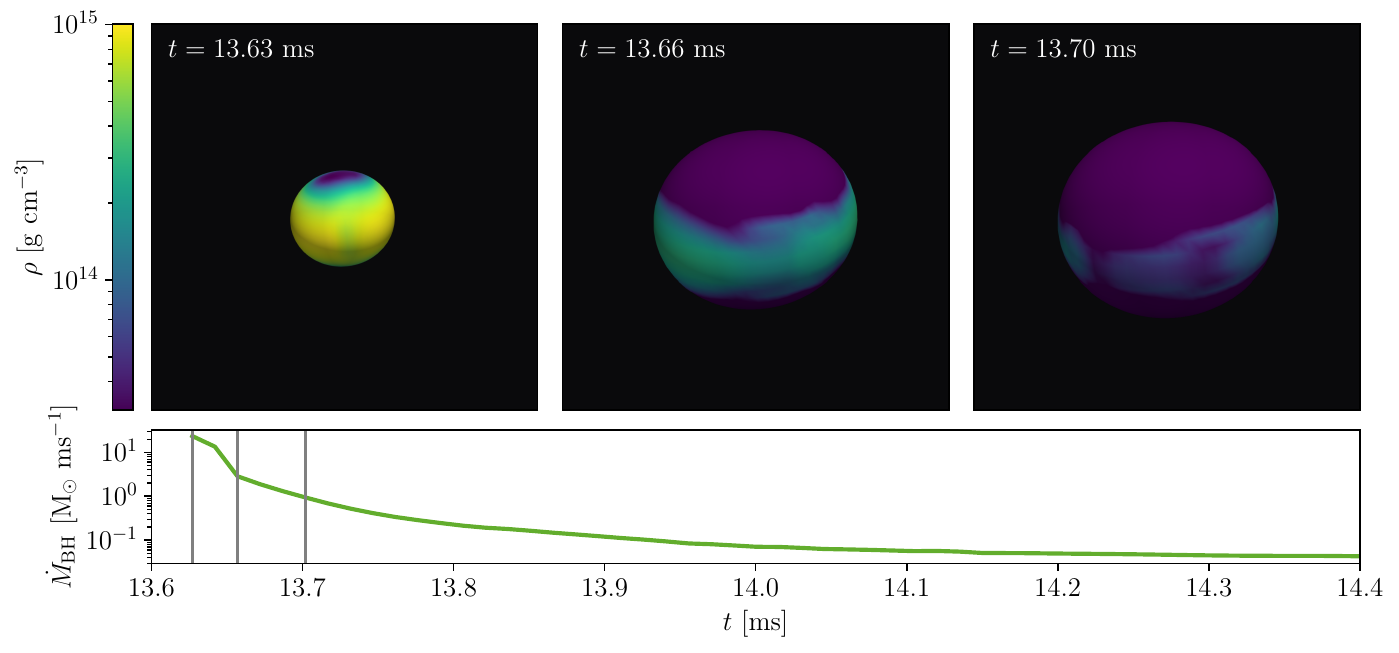}
\caption{Matter-density $\rho$ at the apparent horizon at different times: at the first detection of the horizon, right after its initial rapid expansion, and during slow accretion of matter of the disk. The bottom panel shows the accretion rate (time derivative of the BH mass) over time, with the vertical gray lines representing the times chosen for the three top panels, respectively. We show the data for the equal mass simulation with H4 EOS and $\chi^A=\chi^B=-0.1$. Note that the simulation shown here used a resolution of $h_{L-1}=134$ m.}
\label{fig:AH_panels}
\end{figure*}

\begin{figure}[htpb]
    \centering
    \includegraphics[width=\linewidth]{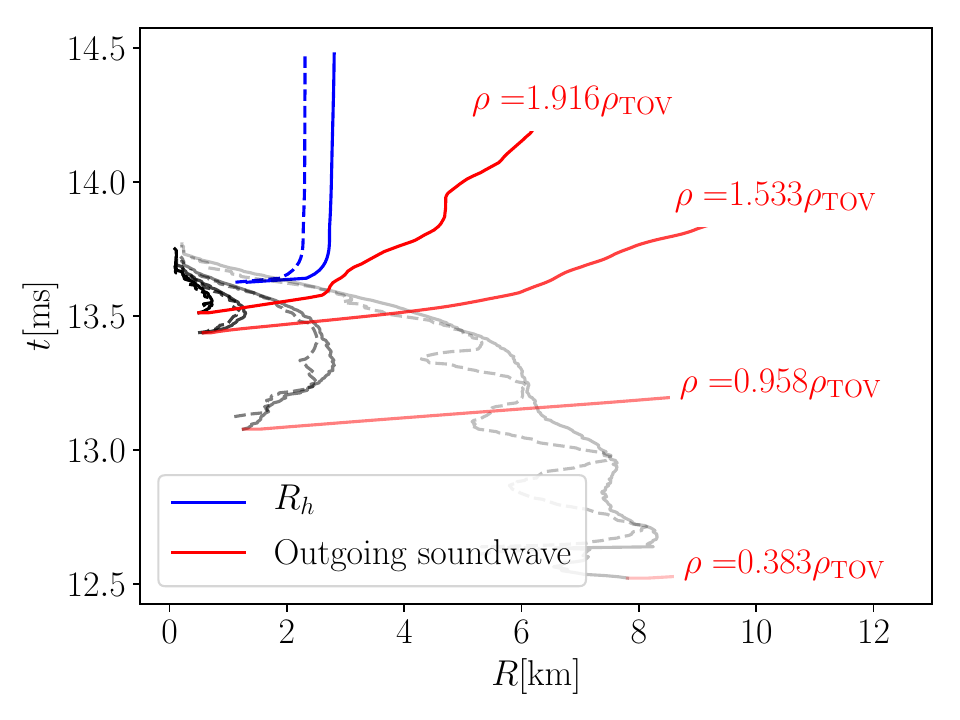}
    \caption{Horizon radius $R_h$ on $x$-$y$ and $x$-$z$ planes. Black solid lines show the outer radius of matter with densities $0.383, 0.958, 1.533, 1.916 \rho_{\mathrm{TOV}}$, from inner to outer radius, respectively. Black dashed lines represent the same quantity but on the $x$-$z$ plane. Blue lines show the apparent horizon, and red lines show the fastest possible trajectory of an outgoing sound wave generated at a certain density. The data is shown for the simulation with H4 EOS, $q=1$, and $\chi^A=\chi^B=-0.1$.}
    \label{fig:Horizon_formation}
\end{figure}

\textit{Apparent-horizon formation:}\footnote{Although our analysis provides information about the BH formation and the maximum densities that can be probed through our simulations, one has to note a few points of caveat. First, our investigations are based on the apparent horizon and not the event horizon, which would be the necessary quantity to investigate causality and if different spacetime regions are causally disconnected from infinity. Second, the apparent-horizon finder can fail in very dynamic spacetimes in which an apparent horizon forms but is significantly deformed. In such a case, it would be possible that an apparent horizon is found at a later time than first formed.}
In the following, we consider the equal mass simulation with H4 EOS and $\chi^A=\chi^B=-0.1$ as an example to investigate which supranuclear densities can be studied through BNS mergers. For this purpose, in Fig.~\ref{fig:AH_panels}, the history of the apparent horizon starting from its formation is depicted, along with the matter-density values right at the apparent-horizon surface~\cite{Thornburg:2006zb}. Right after the formation of the apparent horizon, the density at its surface is around $9\times10^{14}~\mathrm{g~cm^{-3}}$, which is almost two times smaller than the central density of a single star with the maximum mass supported by this EOS, $\rho_{\mathrm{TOV}} \sim 1.6 \times 10^{15} \mathrm{g~cm^{-3}}$. This material is quickly accreted to the BH as the apparent horizon expands. Roughly $30~\mathrm{\mu s}$ after the formation, the BH has almost its final size, and its mass changes only slightly due to accretion (second panel of Fig.~\ref{fig:AH_panels}). $40~\mathrm{\mu s}$ later, the densest part of the disk that was close to the horizon has already fallen inside of it, and slower disk accretion sets in (third panel of Fig.~\ref{fig:AH_panels}), of the order of $\lesssim 1~\mathrm{M_{\odot}~ms^{-1}}$.

\begin{figure*}[htp!]
    \centering
    \hfill
    \begin{minipage}[t]{0.49\linewidth}
        \centering
        \includegraphics[width=\linewidth]{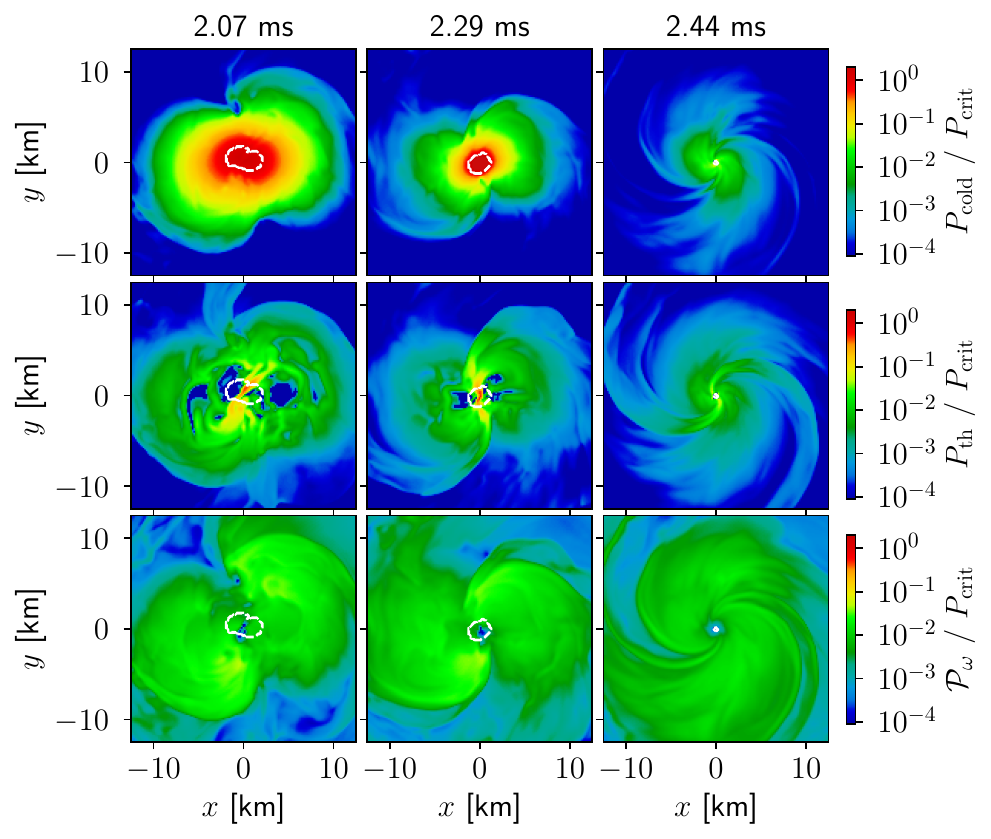}
        {$\chi^A = \chi^B =-0.1$}
    \end{minipage}
    \begin{minipage}[t]{0.49\linewidth}
        \centering
        \includegraphics[width=\linewidth]{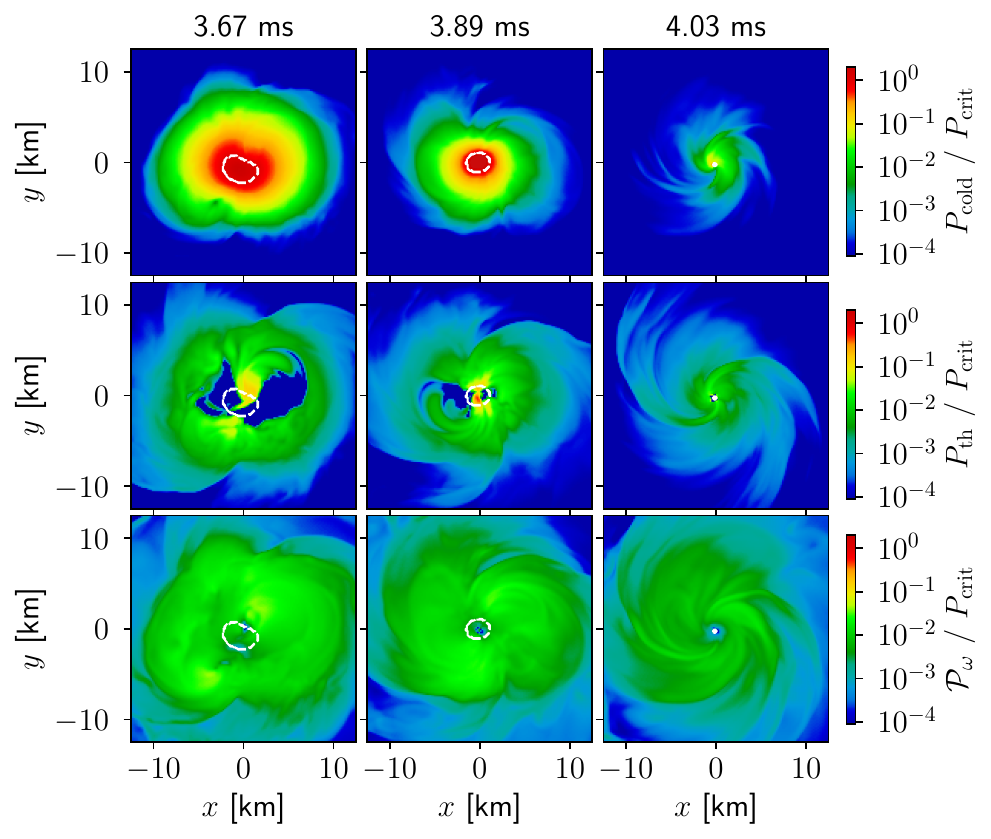}
        {$\chi^A = \chi^B =0.0$}
    \end{minipage}
    \hfill
    \vspace{0.5cm}
    \begin{minipage}[t]{0.49\linewidth}
        \centering
        \includegraphics[width=\linewidth]{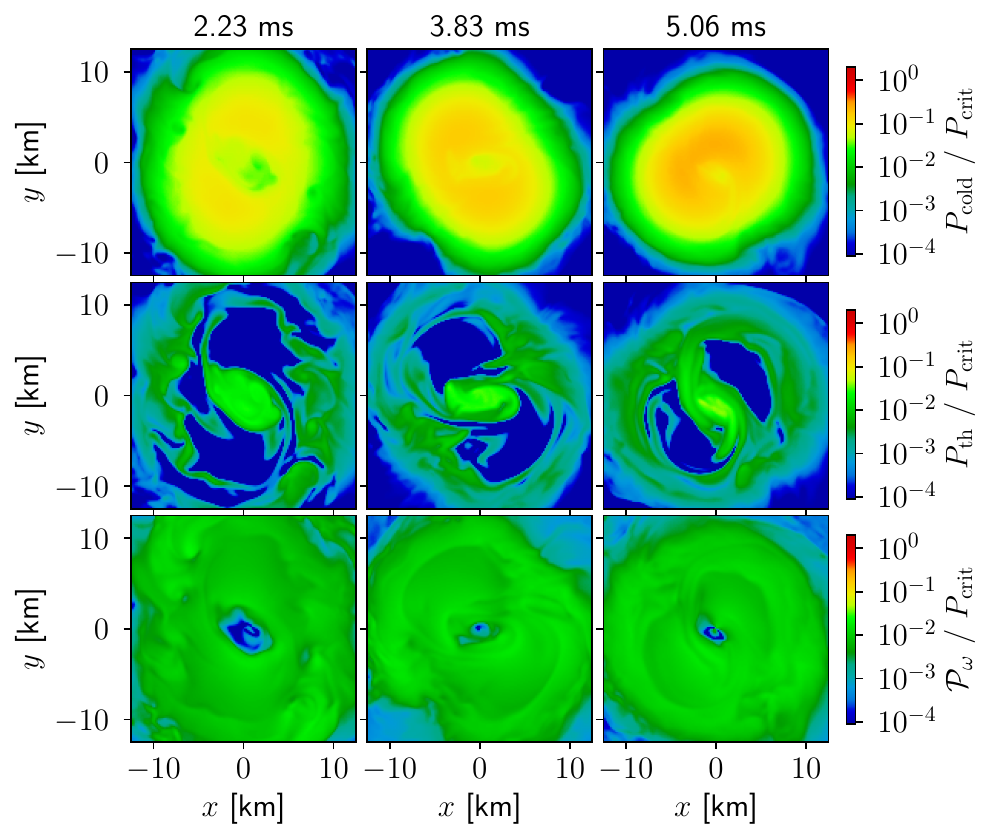}
        {$\chi^A = \chi^B =0.1$}
    \end{minipage}
    \begin{minipage}[t]{0.49\linewidth}
        \centering
        \includegraphics[width=\linewidth]{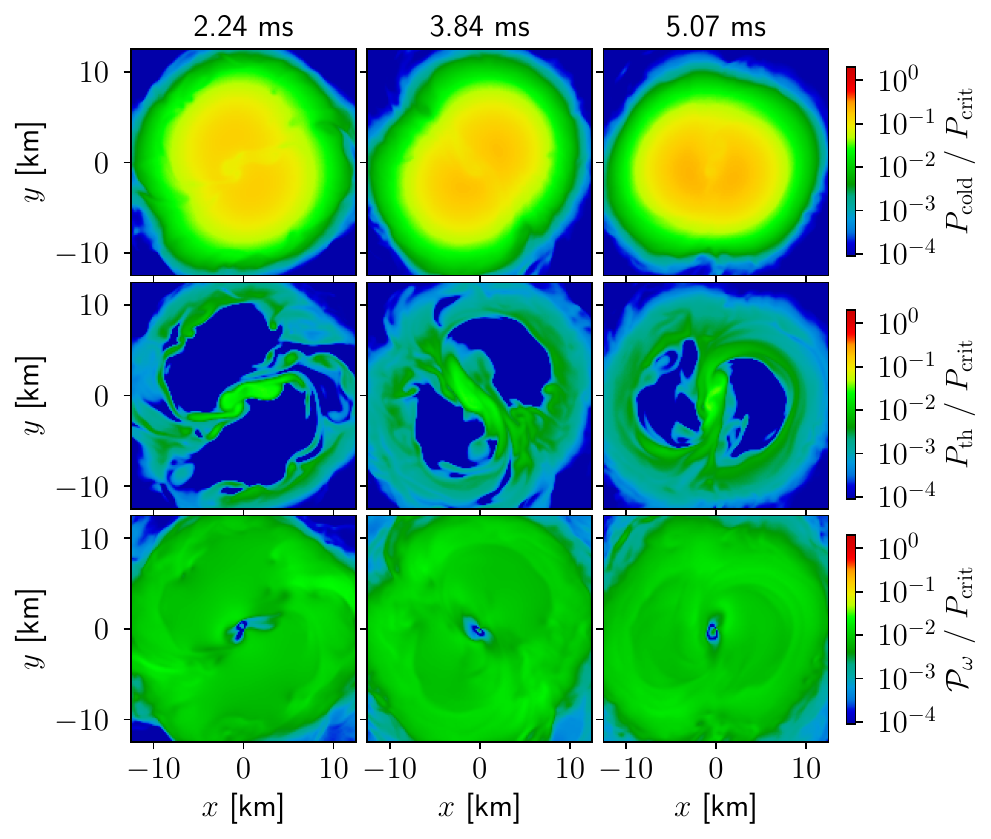}
        {$\chi^A = \chi^B =0.2$}
    \end{minipage}
    \caption{Maps of cold pressure $P_{\rm cold}$, thermal pressure $P_{\rm th}$, and pressure indicator $\mathcal{P}_{\omega}$ associated with the angular momentum for the simulation with H4 EOS and equal mass in the $x$-$y$ plane. The snapshots are extracted at three different times after the merger from refinement levels $5$ and $6$ of the simulations with R3. We compare with respect to the critical pressure of H4 EOS, which is $P_{\rm crit} = 6.32 \times 10^{35}$~${\rm dyn} / {\rm cm}^2$. For the simulations with $t_{\rm coll} < 5$~ms (with $\chi^A = \chi^B =-0.1$ and $\chi^A = \chi^B =0.0$), we show the snapshots around the collapse time and draw the contour of the area with $P_{\rm cold}>P_{\rm crit}$ as a white dashed line.}
    \label{fig:pressure_maps}
\end{figure*}

In Fig.~\ref{fig:Horizon_formation}, we show a time-radius diagram of the remnant's collapse for the H4 simulation with $q=1$ and $\chi^A=\chi^B=-0.1$. Obviously, very high densities can be reached within the apparent horizon, but no information about this spacetime region can escape from the BH. For this reason, the question arises of which densities are probeable before BH formation. For this purpose, we investigate the time shortly before the apparent horizon forms. Given that any shock wave within the remnant will not travel faster than the fastest characteristic velocity of GRHD, we use this speed to compute the maximum density that is probeable through observations before the BH formation. We show in Fig.~\ref{fig:Horizon_formation} contour density lines, where solid lines represent density lines in the $x$-$y$ plane, while dashed lines represent their counterparts in the $x$-$z$ plane. Because of the lack of spherical or axial symmetry, the distance from the remnant of a certain density value is not well defined. For illustration, we set it as the maximum radius at which such densities appear. For each density level shown, we plot the trajectory of an outgoing sound wave emitted at the time and radius the corresponding density first appears in the simulation. Such trajectory is then integrated outward with a radial velocity depending on both time and radius given by ${\rm max} \{ \lambda_r(t,R) \}$, where
\begin{equation}
  \begin{aligned}
    \lambda_r & = -\beta^r + \frac{\alpha}{1-v^2c_s^2} \{ v^r(1-c_s^2) + \\
    & c_s \sqrt{(1-v^2)[\gamma^{rr}(1-v^2c_s^2) -(v^r)^2(1-c_s^2)]} \}
  \end{aligned}
\end{equation}
is the fastest outgoing characteristic velocity of GRHD. Here $c_s$ is the speed of sound, $v^i$ the fluid's 3-velocity, and $\alpha$, $\beta^i$ and $\gamma^{ij}$ the gauge and metric quantities of the 3+1 decomposition of GR.
The choice of the maximum $\lambda_r$ for a given time and radius is aimed to obtain an upper bound of the distance shock waves can travel without being captured into the BH. 
Our analysis shows that densities up to twice the TOV-density $\rho_{\rm TOV}$ can be reached shortly before the formation of the black hole, and the information coming from these regions through shock waves can escape the apparent horizon. This suggests that BNS mergers provide a testbed for probing matter at the most extreme densities. However, we note that while this is theoretically of great interest, the impact of differences above $\rho_{\rm TOV}$ has been shown to be very small~\cite{Ujevic:2023vmo} and will likely not be measurable in the coming decades.

\subsection{Pressure at Collapse}
\label{subsec:pressure}

The formation of the BH depends on the relationship between the strong spacetime curvature and the counteracting pressure supporting the remnant against collapse, e.g., due to repulsive nuclear forces. To get a better understanding of the formation mechanism and to investigate possible spin dependencies, we analyze the behavior of the pressure at the time of collapse. While the dynamics will be determined by the total pressure, we try to assess the importance of individual pressure contributions, i.e., the cold part $P_{\rm cold}$, extracted from the zero-temperature piecewise-polytropic EOS, and the thermal part
\begin{equation}
P_{\rm th} = P - P_{\rm cold}.
\end{equation}

Additionally, we introduce a quantity to investigate the effective pressure counteracting a collapse caused by centrifugal forces. We define this pressure indicator associated with the angular momentum as 
\begin{equation}
     \mathcal{P}_{\omega} = \frac{1}{2} W^2 \rho h \omega^2 r^2,
\end{equation} 
with $W$ being the Lorentz factor, based on the following consideration: For a uniform sphere rotating at uniform angular velocity $\omega$, we have in spherical coordinates
\begin{equation}
    \partial_t (r^2 S_r) + \partial_r \left[ r^2( S_r
    v^r  + P)\right ] = \sqrt{\gamma} \Gamma^{\nu}_{r \mu} T^{\mu}_{\nu},
\end{equation}
neglecting $\phi$ or $\theta$ dependencies, ignoring the gravitational effects, and assuming for the lapse $\alpha = 1$ and the shift $\beta^i = 0$. Under these assumptions, we also can express $S^{i} = \rho h W^2 v^{i}$ and show that
\begin{equation}
    \sqrt{\gamma} \Gamma^{\nu}_{r \mu} T^{\mu}_{\nu} = W^2 \rho h\omega^2 r^3 + 2rP,
\end{equation}
which is divided into a static pressure part and the tidal part that we use to define $\mathcal{P}_{\omega}$.\par

In Fig.~\ref{fig:pressure_maps}, we show maps of the individual pressure components $P_{\rm cold}$, $P_{\rm th}$, and $\mathcal{P}_{\omega}$ for the four simulations with H4 EOS and $q = 1$ at resolution R3. The pressure components are compared with the critical pressure $P_{\rm crit}$, which is the maximum pressure supported by the EOS, i.e., the central pressure of a TOV star with maximum mass $p(\rho_{\rm TOV})$. The simulations differ in the initial spin configuration of the NSs: two with orbit-aligned spins for both stars, one without spin, and one with orbit-antialigned spins for both stars. While we have a prompt collapse for the simulation with orbit-antialigned spins, the zero-spin configuration collapses at $3.96$~ms after the merger, and the configurations with orbit-aligned spins even later (for $\chi^A = \chi^B = 0.1$ no BH is formed during the simulation time of about 50 ms and for $\chi^A = \chi^B = 0.2$ a BH forms only at 13.3~ms after the merger). We present snapshots at different times, cf.~label at the top of the individual plots, measured from the merger of the two stars. 

\begin{figure}[htp!]
    \centering
    \includegraphics[width=0.81\linewidth]{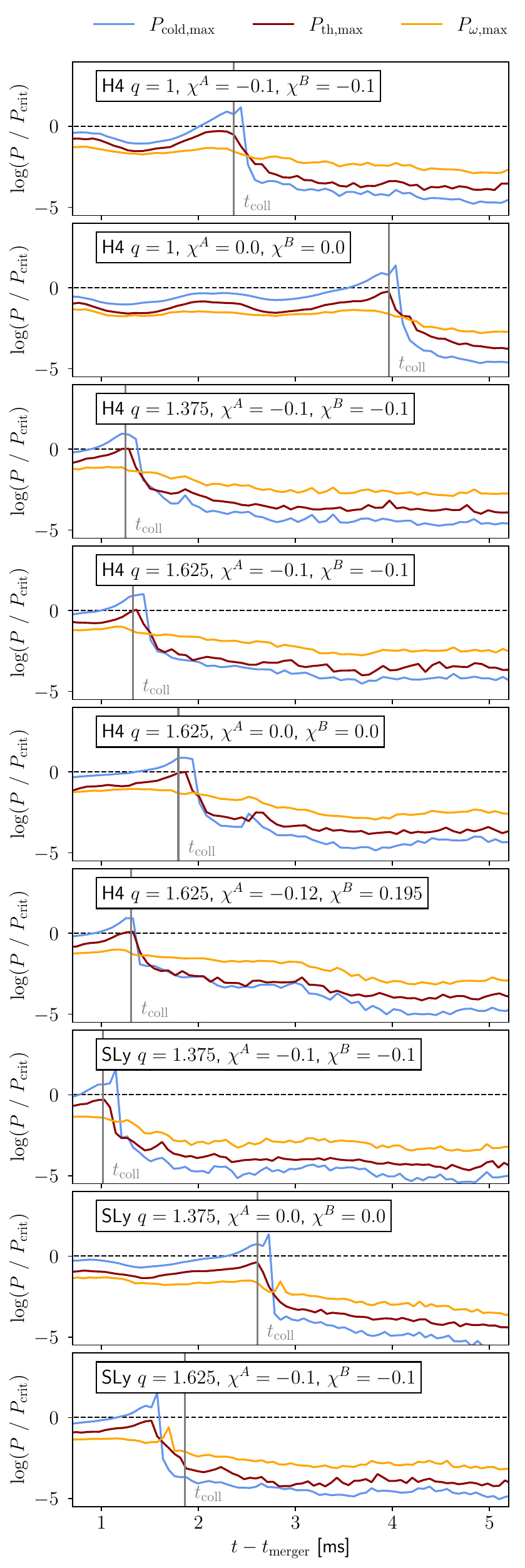}
    \caption{Maxima of cold pressure $P_{\rm cold,max}$, thermal pressure $P_{\rm th,max}$, and pressure indicator associated with the angular momentum $\mathcal{P}_{\omega, \rm max}$ for the R3 simulations with $t_{\rm coll} < 5$~ms. The values are extracted from refinement level $l=6$ in the $x$-$y$ plane. We compare with respect to the critical pressure of the respective EOS and mark $t_{\rm coll}$ as grey vertical lines. }
    \label{fig:pressure1D}
\end{figure}

The total pressure is dominated by the cold component $P_{\rm cold}$. For the two cases with early collapse ($\chi^A = \chi^B = -0.1$ and $\chi^A = \chi^B = 0.0$), $P_{\rm cold}$ exceeds $P_{\rm crit}$. We sketch the region where $P_{\rm cold} > P_{\rm crit}$ with white contour lines in the plots. In the other two cases, $P_{\rm cold}$ is significantly smaller and remains $\lesssim 0.2 P_{\rm crit}$. $P_{\rm th}$ reflects the thermal structure of the remnant. In the simulations with $\chi^A = \chi^B =0.1$ and $\chi^A = \chi^B =0.2$, the two initially cold NS cores are still visible and distinguishable. At the collision surface, $P_{\rm th}$ increases due to shock heating. For the simulations with $\chi^A = \chi^B =-0.1$ and $\chi^A = \chi^B =0.0$, $P_{\rm th}$ actually approaches $P_{\rm crit}$ in the region where also $P_{\rm cold} > P_{\rm crit}$ just before the collapse. The scale of the pressure indicator $\mathcal{P}_{\omega}$, that represents an effective pressure by angular momentum counteracting the collapse, is comparable to $P_{\rm th}$. Yet, in the central region and particularly in the region with $P_{\rm cold} > P_{\rm crit}$, $\mathcal{P}_{\omega}$ remains small and is negligible compared to $P_{\rm cold}$ and $P_{\rm th}$. Overall, it reaches slightly higher values for the orbit-antialigned and non-spinning cases ($\gtrsim 10^{-2} P_{\rm crit}$) than for the systems with orbit-aligned spin ($\lesssim 10^{-2} P_{\rm crit}$). While this seems counterintuitive, since systems with orbit-aligned spins have generally higher angular momentum, it can be explained by the fact that the collision itself is more turbulent in the configurations with zero-spin and orbit-antialigned spins. Thus, higher values for $\mathcal{P}_{\omega}$ can be reached locally. This also explains why in the simulations with $\chi^A = \chi^B =-0.1$ and $\chi^A = \chi^B =0.0$ the cold NS cores are no longer distinguishable in the thermal structure shortly after the merger. The more turbulent collision causes faster mixing of the NS material inside the remnant. \par

We examine the individual pressure components in other simulations with early collapse. In Fig.~\ref{fig:pressure1D}, we show the maxima of $P_{\rm cold}$, $P_{\rm th}$, and $\mathcal{P}_{\omega}$ for the simulations with a collapse time $\lesssim 5$~ms after the merger together with the time of the collapse. As above, the dominant component is $P_{\rm cold}$. Its maxima exceed $P_{\rm crit}$ by a factor of around $10$ to $20$ before the collapses. Also, $P_{\rm th,max}$ approaches the critical pressure. The collapse time corresponds reasonably well to the time at which $P_{\rm th,max}$ reaches its peak and closely approaches $P_{\rm crit}$.\footnote{An exception is the SLy simulation with $q = 1.625$, and $\chi^A = \chi^B =-0.1$. Our hypothesis is that the apparent-horizon finder failed initially to find the apparent horizon due to the larger mass ratio of the system and the potentially very oblate shape.} While the maxima of the cold $P_{\rm cold}$ and thermal $P_{\rm th}$ pressure components subsequently decrease quite drastically, the pressure indicator $\mathcal{P}_{\omega}$ does not seem to be affected too much by the collapse of the remnant.

\subsection{Collapse Time and Spin Effects}

\begin{figure}[t]
    \centering
    \includegraphics[width=\linewidth]{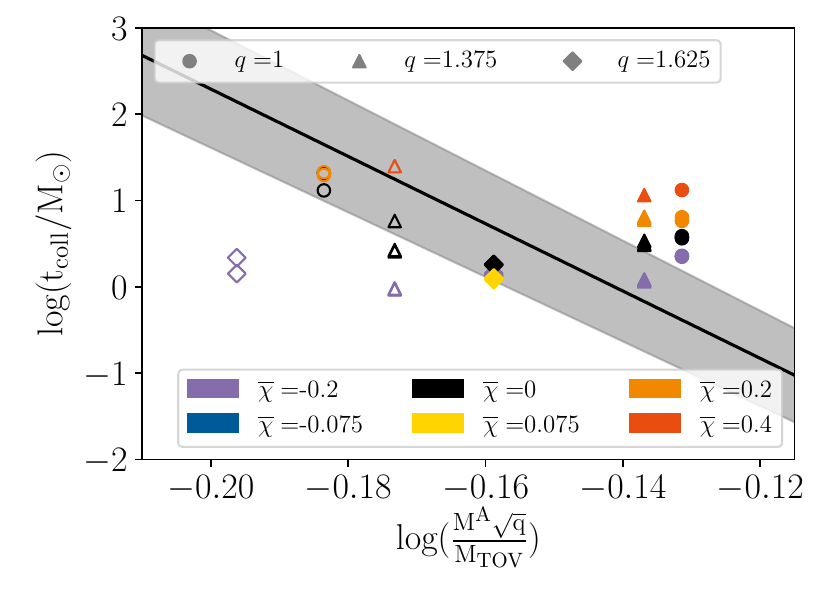}
    \caption{Collapse times of our simulations set compared with the fit of~\cite{Lucca:2019ohp} (solid black line) including the $1\sigma$ confidence interval (gray shaded region). 
    $M_{\rm TOV}$ is the maximum mass of a non-rotating star supported by the EOS. Filled markers represent simulations with H4 EOS, empty ones with SLy.}
    \label{fig:tcoll_fit}
\end{figure}

\begin{table}[t]
\renewcommand{\arraystretch}{1.5}
\setlength{\tabcolsep}{2.2pt}
\caption{Simulations for which we have been able to infer a threshold spin. Columns show the EOS, resolution, mass ratio, the sum of the gravitational masses (which can be interpreted as threshold mass), the threshold spin, and the threshold mass obtained with the fit from ~\cite{Tootle:2021umi}.} 
\label{tab:prompt_collapse_threshold}
\begin{tabular}{c|c|c|c|c|c}

\toprule EOS  & n  & $q$ & $M [M_{\odot}]$ & $\overline{\chi}_{\rm th}$ & $M^{\rm fit}_{\rm th}[M_{\odot}]$  \\ \hline \hline
        H4    & 96  & 1.375 & 3.0 & $-0.114$                & $3.052$  \\
        H4    & 128 & 1.375 & 3.0 & $-0.118$                & $3.050$  \\
        H4    & 160 & 1.375 & 3.0 & $-0.125 \pm 0.01$       & $3.047^{+0.0047}_{-0.0047} \pm 0.03$    \\ \hline
        H4    & 160 & 1.625 & 2.9 & $0.1    \pm 0.1$        & $3.116^{+0.040}_{-0.043} \pm 0.03$      \\ \hline
        SLy   & 160 & 1.375 & 2.8 & $-0.1   \pm 0.1$        & $2.751^{+0.040}_{-0.043} \pm 0.03$      \\ 
 \hline \hline
\end{tabular}
\end{table}

As discussed in the previous section, the presence of NSs' intrinsic spins changes the stability of the remnant and, consequently, the collapse time. In the following, we define the collapse time $t_{\rm coll}$ as the time between the merger time $t_{\rm mrg}$, i.e., the time when the GW amplitude has its maximum, and the time $t_{\rm BH}$ at which an apparent horizon is found first:
\begin{equation}
t_{\rm coll} = t_{\rm BH} - t_{\rm mrg}. 
\end{equation}
As a starting point, we compare the obtained collapse time with the predictions of Ref.~\cite{Lucca:2019ohp}, in which a phenomenological fit for $t_{\rm coll}$ was presented as a function of the gravitational mass of the heavier star and the mass ratio (without including spin effects). Fig.~\ref{fig:tcoll_fit} shows our findings, and we see an overall good agreement between our spinless simulations and the fit. However, for spinning configurations, we observe a larger deviation from the fit. Generally, orbit-antialigned configurations lead to shorter collapse times, while the collapse time is increased for orbit-aligned spin configurations. This observation indicates that for an accurate prediction of the collapse time also spin effects should be included, which, on the other hand, would require a larger number of simulations. \par

Finally, in Tab.~\ref{tab:prompt_collapse_threshold}, we compare our results with the fit for prompt-collapse threshold mass of~\cite{Tootle:2021umi}. While the fit infers the prompt-collapse threshold mass for a given combination of spin and mass ratio, we proceed keeping the latter and the total mass fixed, and inferring the corresponding threshold total spin $\overline{\chi}_{\rm th}$. In both cases, the final result is a set of $M_{\rm th}$, $q_{\rm th}$, and $\overline{\chi}_{\rm th}$, which represent the binary parameters at the prompt-collapse threshold. Following the discussion in \cite{Kolsch:2021lub}, we chose $t_{\rm coll}<2$\,ms as a condition to define a prompt collapse. When this condition is satisfied, the NS cores do not undergo any bounce against each other before collapsing. We proceed with the computation of $\overline{\chi}_{\rm th}$ in the following way: taking $t_{\rm coll}$ for different systems with the same $M$ and $q$, we construct the function $t_{\rm coll}(M,q,\overline{\chi})$ through a linear interpolation of our data points. We solve $t_{\rm coll}(M,q,\overline{\chi}_{\rm th}) = 2$\,ms for $\overline{\chi}_{\rm th}$. The error is then set by the difference between the highest and the lowest resolution. This procedure is applied for the H4 system with $q=1.375$.
In the other cases, the system transits from prompt collapse to no collapse within two successive $\overline{\chi}$ data points; the computation described before is not applicable. In these cases, we set $\overline{\chi}_{\rm th}$ as the average between the last collapsing $\overline{\chi}$ configuration and the next data point, with an error covering the whole total spin bin between the two data points.
The error on $M_{\rm th}$ obtained from the fit is divided in two parts, one derived by our error on $\overline{\chi}_{\rm th}$ and one from the fit performed in Ref.~\cite{Tootle:2021umi}, which we set to the average value reported in that article of $\Delta M_{\rm th}=0.03 M_{\odot}$.

\subsection{Gravitational-Wave Energy and Luminosity}
\label{subsec:GWemission}

\begin{figure}[t]
    \centering
    \includegraphics[width=\linewidth]{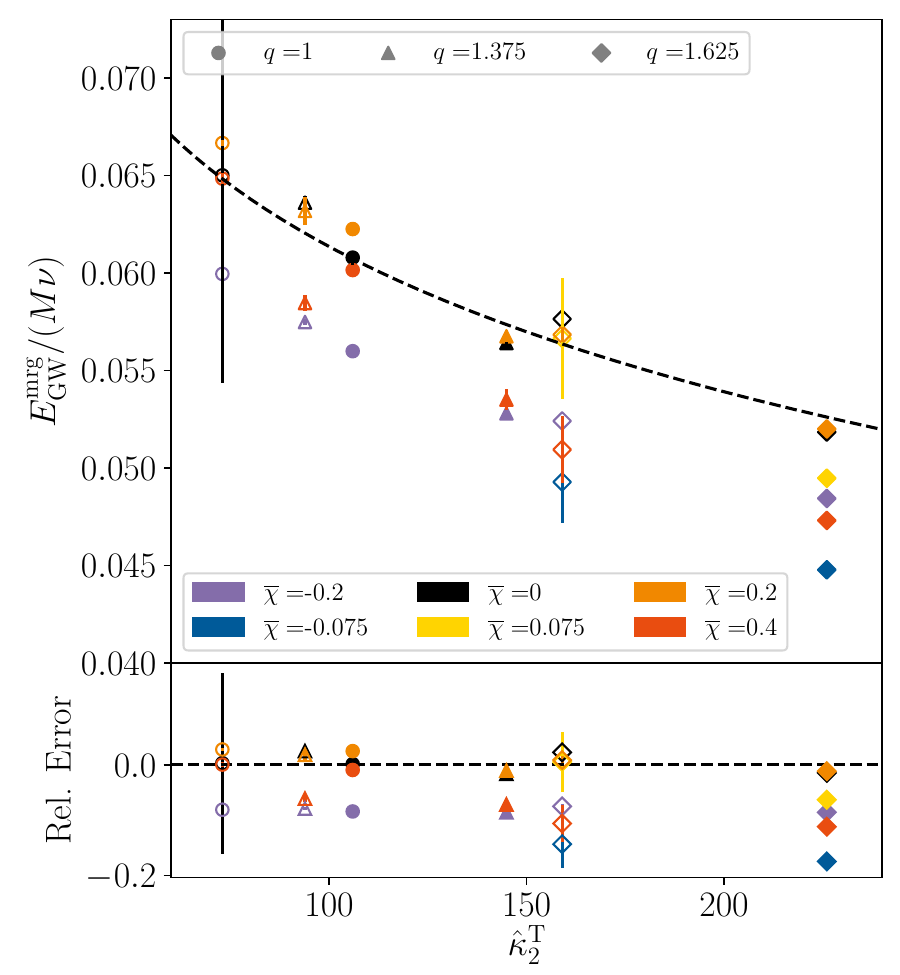}
    \caption{\textbf{Top panel:} Total emitted GW energy as a function of the total quadrupolar tidal polarizability coefficients $\hat{\kappa}_2^T$ for every system. The black dashed line shows the phenomenological relation of \cite{BNS-final-state-T1800417}; in the same article, the variable $\hat{\kappa}_2^T$ is defined. \textbf{Bottom panel:} Relative error with respect to the fitting formula. In both panels, error bars are computed as the difference between the two highest resolutions. Filled markers represent simulations with H4 EOS, empty ones with SLy EOS. }
    \label{fig:GWenergy}
\end{figure}

\begin{figure}[t]
    \centering
    \includegraphics[width=\linewidth]{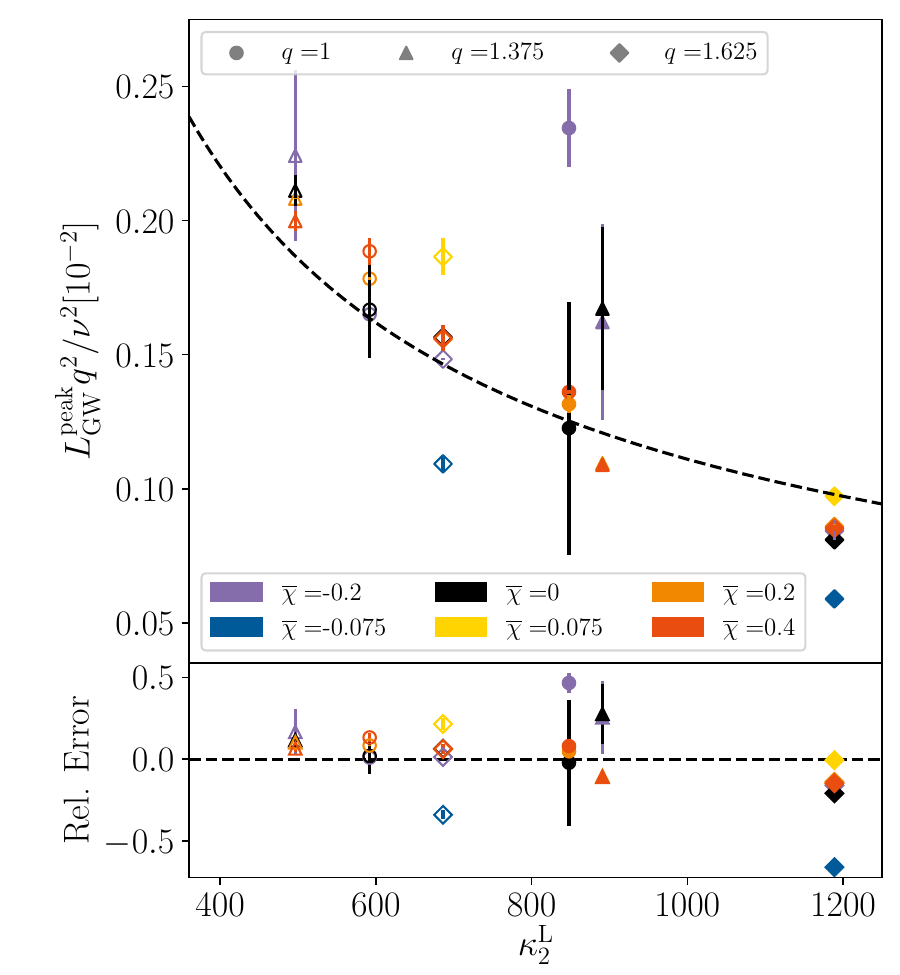}
    \caption{Same as Fig. \ref{fig:GWenergy} but for GW luminosity. Note the fit is now performed as a function of a different tidal polarizability parameter. The definition of $\kappa^L_2$ can be found in \cite{Zappa:2017xba}. Filled markers represent simulations with H4 EOS, empty ones with SLy EOS.}
    \label{fig:GWluminosity}
\end{figure}

In the following, we also investigate the spin influence on the emitted GW energy and on the GW peak luminosity. We compute the total energy radiated by GWs at the merger as
\begin{equation}
     E_{\rm GW}(t_{\rm mrg}) = M^A + M^B - M_{\rm ADM}(t_0) + E_{\rm GW}^{\rm sim}(t_{\rm mrg}),
\end{equation}
with $E_{\rm GW}^{\rm sim}(t_{\rm mrg})$ being the GW energy radiated during the simulation up to the merger $t_{\rm mrg}$, i.e.,
\begin{equation}
    E_{\rm GW}^{\rm sim}(t_{\rm mrg}) = \frac{r^2_{\rm ext}}{16 \pi} \int d\Omega \int_{t_0}^{t_{\rm mrg}} dt \left | \int_{t_0}^t dt' \Psi_4 \right |^2
\end{equation}
where $r_{\rm ext} \simeq 1350$\,km is the radius at which the GW signal is extracted, and $\Psi_4$ is the Weyl scalar. \par 

We present our results in Fig.~\ref{fig:GWenergy}, where we compare the GW total radiated energy up to the merger with the fit of \cite{BNS-final-state-T1800417}. Also, in this case, the fits are derived purely based on non-spinning simulations. We observe that orbit-aligned spin systems with $\overline{\chi}=\{ 0, 0.2 \}$ follow the fitting curve, while systems with higher or negative total spin systematically emit less with respect to their counterparts. More importantly, if we look at the spin-antialigned systems, one can see that for $\overline{\chi}=-0.075$ the emitted energy reaches systematically lower values with respect to systems with $\overline{\chi}=-0.2$.  In fact, systems for which we have a spin-antialigned setup lead to quite different simulation outcomes, cf.~tables in App~\ref{app:tables}, although in both case $\tilde{\chi}$ = 0. This suggests that a more complicated spin dependence than one based on either $\tilde{\chi}$ or $\overline{\chi}$ has to be considered for accurate modeling of the systems. 

In Fig.~\ref{fig:GWluminosity}, we show the GW peak luminosity, defined as $L_{\rm GW}^{\rm peak} = \rm max_t \{ dE_{\rm GW}/dt \}$. In this case, our data are compared with the fit of Ref.~\cite{Zappa:2017xba}, where the peak luminosity is recalled by the mass ratio $q$ divided by the symmetric mass ratio $\nu = M^{\rm A} M^{\rm B}/ M^2$. In contrast to the previous case, we do not observe any clear evidence of a systematic spin effect.

\subsection{Disk Mass Estimates}

\begin{figure}[t]
    \centering
    \includegraphics[width=\linewidth]{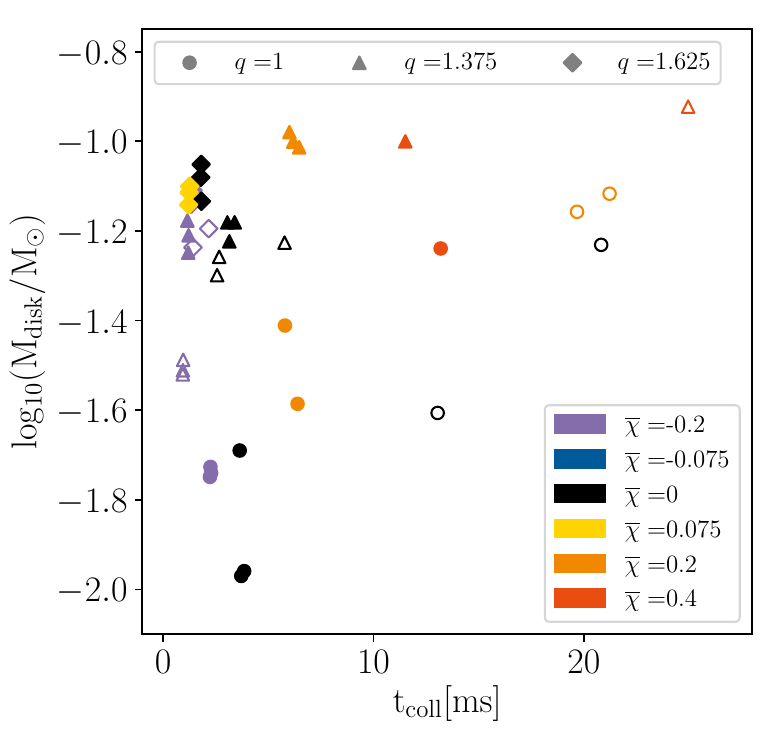}
    \caption{Disk mass and collapse time for all the simulations where we observe a collapse into a BH. Different mass ratios are shown with different markers and different spin configurations with different colors. Empty markers represent simulations with SLy EOS and filled ones with H4.}
    \label{fig:Disk_mass}
\end{figure}

Finalizing our discussion about the simulation results, we present the measured disk masses for all our setups that undergo a collapse (employing the highest resolution) as a function of the collapse time in Fig.~\ref{fig:Disk_mass}. We compute the mass of the disk as the integral of conserved bound baryonic mass density over the region outside the apparent horizon after the remnant's collapse. While we do not find a clear functional behavior between the collapse time and the disk mass, one can conclude from Fig.~\ref{fig:Disk_mass} that setups in which the collapse is delayed generally produce larger disk masses. This is clearly visible in our equal-mass setups and can be explained by the observation that sufficient time is needed to redistribute the material in the formed merger remnant. The figure also suggests that setups with larger mass ratios (see different markers in Fig.~\ref{fig:Disk_mass}) are able to produce larger disk masses, even if the collapse time is short, e.g.,~\cite{Dietrich:2016hky,Bernuzzi:2020txg}. This larger disk mass is due to the large amount of matter tidally shed by the secondary component during the merger process.  
Finally, our sample of numerical-relativity simulations also allows us to (i) see that the collapse time increases with orbit-aligned spin, as described before, but also that (ii) the disk mass generally tends to increase if the individual stars had an intrinsic spin aligned to the orbital angular momentum. Both effects are due to the larger angular momentum of the remnant at the time of the merger, which causes a delayed BH formation and an increased material transport to the outward regions of the formed remnant.

\section{Summary}
\label{sec:conclusion}
In this work, we simulate a total of $28$~BNS systems close to the prompt-collapse threshold, with two different EOSs, three mass ratios, and four spin configurations, plus four spin-antialigned configurations for the highest mass ratio. For each setup, we run three different resolutions in order to establish the robustness of our results and to perform error estimation. The simulation results will be released as part of the CoRe database~\cite{Dietrich:2018phi,Gonzalez:2022mgo} in the near future. 

First, we investigate the collapse dynamics of a reference setup that undergoes a BH collapse, even though not promptly. This analysis suggests that densities above $\rho_{\mathrm{TOV}}$ can, in principle, be investigated since, in case of non-prompt collapse, information has enough time to travel from such dense regions to the outside before the horizon forms. However, in practice, probing properties from the aforementioned regions could be very difficult, as shown in previous studies~\cite{Ujevic:2023vmo}. 
Investigating the pressure at the collapse for a few reference simulations undergoing gravitational collapse shows that near the collapse, the pressure is fully dominated by the cold component. In all simulated cases, the thermal component grows near the critical value right at the collapse time. We define the pressure indicator $P_{\omega}$, which has the scope of quantifying the outward pointing pressure due to the star's rotation. We find that  $P_{\omega}$ always remains smaller than the threshold pressure, especially at the collapse. 

Based on the different collapse times of systems with different spins, we infer the threshold parameters $M_{\rm th}$, $q_{\rm th}$ and $\overline{\chi}_{\rm th}$, which separate the prompt-collapse region from the rest of the parameter space. We compare our results with~\cite{Tootle:2021umi} and find, except for one case, agreement between our results and the proposed phenomenological relation of~\cite{Tootle:2021umi}. 

We compare collapse times with the fit performed in \cite{Lucca:2019ohp}, which does not include spin effects. Our findings for non-spinning systems are in line with the fit. However, spinning simulations show a deviation. This suggests the necessity of including spin effects in the fit of $t_{\rm coll}$ to obtain a reliable estimate of the collapse time. Similarly, we find that the energy carried away by GWs is in line with the fit performed in \cite{Zappa:2017xba}, but spinning systems show the necessity of a more detailed fitting function. The picture is different for the maximum GW luminosity, where our results are in line with the fit of \cite{Zappa:2017xba}, but do not show any clear trend related to the spin. 

The effect of NS spins on the disk mass is also visible. The longer $t_{\rm coll}$ of systems with high orbit-aligned spins is reflected in a higher mass of the disk, with the orbit-antialigned spin configurations giving the lowest disk masses. 

Finally, it is worth pointing out that while the setups with $\overline{\chi}=\{-0.075,~0,~0.075\}$ should be very close to each other, the first one significantly differs from the others in every analysis we performed. If corroborated by more data, this could suggest the need for a more complicated description of the spin effects than one based only on the total spin $\overline{\chi}$ or the mass-averaged spin $\tilde{\chi}=0$.

\section{Acknowledgments}

We thank Bernd Br\"ugmann and Maximilian K\"olsch for helpful discussions that stimulated the work on this project. Our thanks also go to Henrik Rose for reviewing this manuscript. 
The project was supported by the European Union (ERC, SMArt, 101076369). Views and opinions expressed are those of the authors only and do not necessarily reflect those of the European Union or the European Research Council. Neither the European Union nor the granting authority can be held responsible for them.
A.\ N.\ and I.\ M.\ gratefully acknowledge the support of Deutsche Forschungsgemeinschaft (DFG) through Project No.~504148597. 

Further, we gratefully acknowledge the Gauss Centre for Supercomputing e.V.\ for funding this project by providing computing time on the GCS Supercomputer SuperMUC-NG at Leibniz Supercomputing Centre [project pn29ba], the national supercomputer HPE Apollo Hawk at the High Performance Computing (HPC) Center Stuttgart (HLRS) under the grant number GWanalysis/44189, as well as the Resource Allocation Board providing computing time on the supercomputer Lise/Emmy as part of the NHR infrastructure [project bbp00049] and the SDumont supercomputer of the National Laboratory for Scientific Computing (LNCC/MCTI, Brazil) [project 237565].

We thank Berlin Public Transport Company (BVG) for designing the color palette used throughout this article.

\bibliography{refs}

\appendix

\section{Summary Tables}
\label{app:tables}

We summarize our findings in Tabs.~\ref{tab:H4_aligned}, \ref{tab:SLy_aligned}, and \ref{tab:anti-aligned}. 

The remnant mass $M_{\rm BH}$ and dimensionless spin parameter $\chi_{\rm BH} = J_{\rm BH} / M_{\rm BH}^2$ are calculated from its apparent horizon with

\begin{align}
    J_{\rm BH} &= \frac{1}{8 \pi } \oint \varphi^l s^m K_{lm} dA, \\
    M_{\rm BH} &= \sqrt{\frac{A_{\rm BH}}{16 \pi} + \frac{4 \pi J_{\rm BH}^2}{A_{\rm BH}}},
\end{align}

\noindent where $A_{\rm BH}$ is the surface area of the horizon, $s^m$ the unit outward pointing normal on the horizon, $\varphi^l$ a Killing vector field on the horizon, and $K_{lm}$ the extrinsic curvature on the spatial hypersurface. \par 

\begin{table*}[htp]
\caption{Properties of the individual stars and BNS simulations for the H4 EOS. The columns contain, respectively, EOS, employed mass ratio $q=M^A/M^B\geq 1$, total gravitational mass,  quadrupolar tidal polarizability coefficients ~\cite{Damour:2009wj} of the single stars and their reduced spin parameters, ADM mass and angular momentum of the system, the initial dimensionless GW frequency $M\omega^0 _{2,2}$ of the (2,2)-mode, resolution in the number of points along one direction, time of the merger, remnant's collapse time, mass, and spin of the resulting BH, mass of the disk, total energy radiated via GWs, and GW luminosity at merger.} 
\label{tab:H4_aligned}
\begin{tabular}{ccccccccccccccccccc}
\toprule 
EOS & $q$ & $M$ & $\kappa_2^A$ & $\kappa_2^B$ & $\chi^A$ & $\chi^B$ & $\overline{\chi}$ &$M_{\rm ADM}$ & $J_{\rm ADM}$ & $ M \omega^0_{2,2}$ & $n$ & $t_{\rm mrg}$ &  $t_{\rm coll}$ & $M_{\rm BH}$ & $\chi_{\rm BH}$ & $M_{\rm disk}$ & $e^{\rm mrg}_{\rm GW}$ & $L^{\rm mrg}_{\rm GW}$ \\
& & $[M_{\odot}]$ & & & & & & $[M_{\odot}]$ & $[M^2_{\odot}]$ & $[10^{-2}]$ & & $[\rm ms]$ &  $[\rm ms]$ & $[M{\odot}]$ & & $[10^{-2} M_{\odot}]$ & $[10^{-2} M_{\odot}]$ & $[10^{-3}]$ \\
\hline \hline
           H4 & 1.00 & 3.0 & 53 & 53 & -0.1 & -0.1  & -0.2 & 2.97155 & 8.39379 & 5.097 & 96  & 10.72 & 2.29 & 2.88 & 0.76 & 1.82 & 5.548  & 0.727 \\
              & &      &     &       &      &       &      &         &         &       & 128 & 10.97 & 2.24 & 2.88 & 0.75 & 1.78 & 5.590  & 0.749 \\
              & &      &     &       &      &       &      &         &         &       & 160 & 10.93 & 2.36 & 2.89 & 0.78 & 1.88 & 5.599  & 0.757 \\
              & &      &     &       & 0.0  & 0.0   &  0.0 & 2.97035 & 8.80087 & 5.089 & 96 & 12.29 & 3.71  & 2.85 & 0.76 & 2.04 & 5.973  & 0.858 \\
              & &      &     &       &      &       &      &         &         &       & 128 & 12.45 & 3.82 & 2.86 & 0.78 & 1.07 & 6.043  & 0.923 \\
              & &      &     &       &      &       &      &         &         &       & 160 & 12.53 & 3.96 & 2.87 & 0.76 & 1.10 & 6.079  & 0.957 \\
              & &      &     &       & 0.1  & 0.1   & 0.2  & 2.97142 & 9.22215 & 5.084 & 96  & 13.57 & 6.01 & 2.80 & 0.74 & 3.88 & 6.184  & 1.067 \\
              & &      &     &       &      &       &      &         &         &       & 128 & 13.73 & 6.47 & 2.84 & 0.77 & 2.60 & 6.208  & 1.081 \\
              & &      &     &       &      &       &      &         &         &       & 160 & 13.82 & ---  &  --- &  --- & ---  & 6.225  & 1.085 \\
              & &      &     &       & 0.2  & 0.2   & 0.4  & 2.97441 & 9.62658 & 5.076 & 96  & 14.23 & ---  &  --- &  --- & ---  & 5.979  & 1.161 \\
              & &      &     &       &      &       &      &         &         &       & 128 & 14.40 & ---  &  --- &  --- & ---  & 6.019  & 1.199 \\
              & &      &     &       &      &       &      &         &         &       & 160 & 14.46 & 13.3 & 2.78 & 0.77 & 5.76 & 6.015  & 1.195 \\

              & 1.375 & 3.0 & 26 & 89 & -0.1 & -0.1  & -0.2 & 2.97244 & 8.15903 & 5.097 & 96  & 10.29 & 1.33  & 2.80 & 0.71 & 6.17 & 5.242  & 0.907 \\
              & &           &    &    &      &       &      &         &         &       & 128 & 9.921 & 1.84  & 2.81 & 0.72 & 5.64 & 5.259  & 0.890 \\
              & &           &    &    &      &       &      &         &         &       & 160 & 10.56 & 1.24  & 2.81 & 0.72 & 6.66 & 5.280  & 0.913 \\
              & &           &    &    & 0.0  & 0.0   & 0.0  & 2.97108 & 8.58404 & 5.087 & 96  & 12.02 & 4.51  & 2.75 & 0.71 & 6.59 & 5.601  & 0.924 \\
              & &           &    &    &      &       &      &         &         &       & 128 & 12.33 & 2.95  & 2.78 & 0.73 & 5.98 & 5.636  & 0.955 \\
              & &           &    &    &      &       &      &         &         &       & 160 & 12.23 & 4.38  & 2.78 & 0.73 & 6.60 & 5.641  & 0.955 \\
              & &           &    &    & 0.1  & 0.1   & 0.2  & 2.97229 & 9.02591 & 5.083 &  96 & 13.19 & 16.2  & 2.64 & 0.69 & 9.99 & 5.606  & 0.948 \\
              & &           &    &    &      &       &      &         &         &       & 128 & 13.38 & 7.21  & 2.71 & 0.72 & 10.5 & 5.646  & 0.970 \\
              & &           &    &    &      &       &      &         &         &       & 160 & 13.49 & 6.54  & 2.73 & 0.73 & 9.70 & 5.676  & 1.001 \\
              & &           &    &    & 0.2  & 0.2   & 0.4  & 2.97566 & 9.44874 & 5.073 & 96  & 13.78 & ---   & ---  &  --- & ---  & 5.246  & 0.957 \\
              & &           &    &    &      &       &      &         &         &       & 128 & 13.98 & ---   & ---  &  --- & ---  & 5.296  & 0.994 \\
              & &           &    &    &      &       &      &         &         &       & 160 & 14.11 & 11.6  & 2.69 & 0.72 & 9.99 & 5.351  & 1.037 \\
              
              & 1.625 & 2.9 & 23 & 135 & -0.1 & -0.1  & -0.2 & 2.87515 & 7.42804 & 4.863 &  96 & 10.98 & 1.52  & 2.67 & 0.65 & 7.77 & 4.784  & 0.687 \\
              & &           &    &     &      &       &      &         &         &       & 128 & 11.23 & 1.45  & 2.69 & 0.67 & 7.24 & 4.829  & 0.735 \\
              & &           &    &     &      &       &      &         &         &       & 160 & 11.42 & 1.32  & 2.69 & 0.67 & 7.76 & 4.844  & 0.747 \\
              & &           &    &     & 0.0  & 0.0   & 0.0  & 2.87372 & 7.85166 & 4.854 & 96  & 12.84 & 1.99  & 2.64 & 0.67 & 8.88 & 5.117  & 0.681 \\
              & &           &    &     &      &       &      &         &         &       & 128 & 12.99 & 1.76  & 2.67 & 0.69 & 7.35 & 5.177  & 0.751 \\
              & &           &    &     &      &       &      &         &         &       & 160 & 13.03 & 1.79  & 2.67 & 0.69 & 8.31 & 5.184  & 0.757 \\
              & &           &    &     & 0.1  & 0.1   & 0.2  & 2.87500 & 8.29095 & 4.849 & 96  & 14.19 & --    & ---  &  ---  & ---  & 5.139 & 0.781 \\
              & &           &    &     &      &       &      &         &         &       & 128 & 14.45 & ---   & ---  &  ---  & ---  & 5.173 & 0.796 \\
              & &           &    &     &      &       &      &         &         &       & 160 & 14.60 & ---   & ---  &  ---  & ---  & 5.200 & 0.820 \\
              & &           &    &     & 0.2  & 0.2   & 0.4  & 2.87855 & 8.71140 & 4.840 & 96  & 14.88 & ---   & ---  &  ---  & ---  & 4.689 & 0.778 \\
              & &           &    &     &      &       &      &         &         &       & 128 & 15.03 & ---   & ---  &  ---  & ---  & 4.713 & 0.797 \\
              & &           &    &     &      &       &      &         &         &       & 160 &  15.14 & --   & ---  &  ---  & ---  & 4.731 & 0.810 \\
 \hline \hline
\end{tabular}
\end{table*}

\begin{table*}[htp]
\caption{Same as Table~\ref{tab:H4_aligned} for SLy EOS.}  
\label{tab:SLy_aligned}
\begin{tabular}{cccccccccccccccccccc}
    \toprule
EOS & $q$ & $M$ & $\kappa_2^A$ & $\kappa_2^B$ & $\chi^A$ & $\chi^B$ & $\overline{\chi}$ &  $M_{\rm ADM}$ & $J_{\rm ADM}$ & $M\omega^0_{2,2}$ & $n$ & $t_{\rm mrg}$ &  $t_{\rm coll}$ & $M_{\rm BH}$ & $\chi_{\rm BH}$ & $M_{\rm disk}$ & $e^{\rm mrg}_{\rm GW}$ & $L^{\rm mrg}_{\rm GW}$ \\
& & $[M_{\odot}]$ & & & & & & $[M_{\odot}]$ & $[M^2_{\odot}]$ & $[10^{-2}]$ & & $[\rm ms]$ &  $[\rm ms]$ & $[M{\odot}]$ & & $[10^{-2} M_{\odot}]$ & $[10^{-2} M_{\odot}]$ & $[10^{-3}]$ \\
\hline \hline
          SLy & 1.00 & 2.7 &  37 & 37  & -0.1 & -0.1 &-0.2  & 2.67658 & 7.06225 & 4.400 & 96  & 17.68 & ---     & ---   & ---   & ---     & 5.917 & 0.995 \\
              &      &     &     &     &      &      &      &         &         &       & 128 & 17.93 & ---     & ---   & ---   & ---     & 5.958 & 1.013 \\
              & &          &     &     &      &      &      &         &         &       & 160 & 18.07 & ---     & ---   & ---   & ---     & 5.995 & 1.046 \\
              & &          &     &     & 0.0  & 0.0  & 0.0  & 2.67551 & 7.37877 & 4.394 & 96  & 19.37 & ---     & ---   & ---   & ---     & 6.383 & 1.162 \\
              & &          &     &     &      &      &      &         &         &       & 128 & 18.62 & 13.12   & 2.51  & 0.65  & 2.48    & 5.438 & 0.245 \\
              & &          &     &     &      &      &      &         &         &       & 160 & 19.78 & 20.88   & 2.45  & 0.63  & 5.87    & 6.501 & 1.274 \\
              & &          &     &     & 0.1  & 0.1  & 0.2  & 2.67649 & 7.70736 & 4.390 & 96  & 20.69 & ---     & ---   & ---   & ---     & 6.541 & 1.336 \\
              & &          &     &     &      &      &      &         &         &       & 128 & 20.98 & 19.76   & 2.42  & 0.62  & 6.96    & 6.651 & 1.460 \\
              & &          &     &     &      &      &      &         &         &       & 160 & 21.11 & 21.29   & 2.41  & 0.61  & 7.64    & 6.666 & 1.467 \\
              & &          &     &     & 0.2  & 0.2  & 0.4  & 2.67911 & 8.01174 & 4.382 & 96  & 21.27 & ---     & ---   & ---   & ---     & 6.364 & 1.444 \\
              & &          &     &     &      &      &      &         &         &       & 128 & 21.56 & ---     & ---   & ---   & ---     & 6.454 & 1.557 \\
              & &          &     &     &      &      &      &         &         &       & 160 & 21.70 & ---     & ---   & ---   & ---     & 6.484 & 1.594 \\
              
              & 1.375 & 2.8 & 15  & 49 & -0.1 & -0.1 & -0.2 & 2.77607 & 7.25651 & 4.629 & 96  & 15.73 & 1.00  & 2.67  & 0.73  & 3.02    & 5.719 & 1.494 \\
              & &           &     &    &      &      &      &         &         &       & 128 & 15.93 & 0.99  & 2.67  & 0.73  & 3.08    & 5.735 & 1.509 \\
              & &           &     &    &      &      &      &         &         &       & 160 & 16.01 & 1.01  & 2.67  & 0.73  & 3.25    & 5.749 & 1.518 \\
              & &           &     &    & 0.0  & 0.0  & 0.0  & 2.77447 & 7.64456 & 4.621 & 96  & 17.44 & 2.81  & 2.61  & 0.72  & 5.52    & 6.270 & 1.650 \\
              & &           &     &    &      &      &      &         &         &       & 128 & 17.65 & 5.83  & 2.59  & 0.70  & 5.94    & 6.328 & 1.755 \\
              & &           &     &    &      &      &      &         &         &       & 160 & 17.78 & 2.60  & 2.62  & 0.73  & 5.03    & 6.361 & 1.798 \\
              & &           &     &    & 0.1  & 0.1  & 0.2  & 2.77592 & 8.04818 & 4.616 & 96  & 18.81 & ---   & ---   & ---   & ---     & 6.262 & 1.780 \\
              & &           &     &    &      &      &      &         &         &       & 128 & 19.04 & ---   & ---   & ---   & ---     & 6.391 & 1.952 \\
              & &           &     &    &      &      &      &         &         &       & 160 & 19.10 & ---   & ---   & ---   & ---     & 6.318 & 1.850 \\
              & &           &     &    & 0.2  & 0.2  & 0.4  & 2.77977 & 8.41755 & 4.604 & 96  & 19.28 & ---   & ---   & ---   & ---     & 6.116 & 1.783 \\
              & &           &     &    &      &      &      &         &         &       & 128 & 19.29 & 25.02 & 2.45 & 0.64 & 11.9      & 5.807 & 1.778 \\
              & &           &     &    &      &      &      &         &         &       & 160 & 19.43 & ---   & ---   & ---   & ---     & 5.846 & 1.813 \\
            
              & 1.625 & 2.7 & 14  & 77 & -0.1 & -0.1 & -0.2 & 2.67873 & 6.55937 & 4.401 & 96  & 17.42 & ---    & ---   & ---   & ---     & 5.150 & 1.191 \\
              & &           &     &    &      &      &      &         &         &       & 128 & 17.71 & 2.24   & 2.50  & 0.64  & 6.39    & 5.220 & 1.278 \\
              & &           &     &    &      &      &      &         &         &       & 160 & 17.85 & 1.86   & 2.51  & 0.65  & 5.80    & 5.241 & 1.288 \\
              & &           &     &    & 0.0  & 0.0  & 0.0  & 2.67690 & 6.96443 & 4.392 & 96  & 19.87 & ---    & ---   & ---   & ---     & 5.828 & 1.491 \\
              & &           &     &    &      &      &      &         &         &       & 128 & 19.79 & ---    & ---   & ---   & ---     & 5.747 & 1.383 \\
              & &           &     &    &      &      &      &         &         &       & 160 & 19.91 & ---    & ---   & ---   & ---     & 5.764 & 1.413 \\
              & &           &     &    & 0.1  & 0.1  & 0.2  & 2.67859 & 7.38489 & 4.387 & 96  & 20.94 & ---    & ---   & ---   & ---     & 5.600 & 1.332 \\
              & &           &     &    &      &      &      &         &         &       & 128 & 21.20 & ---    & ---   & ---   & ---     & 5.644 & 1.390 \\
              & &           &     &    &      &      &      &         &         &       & 160 & 21.39 & ---    & ---   & ---   & ---     & 5.684 & 1.426 \\
              & &           &     &    & 0.2  & 0.2  & 0.4  & 2.68302 & 7.76762 & 4.373 & 96  & 21.41 & ---    & ---   & ---   & ---     & 5.204 & 1.378 \\
              & &           &     &    &      &      &      &         &         &       & 128 & 21.69 & ---    & ---   & ---   & ---     & 5.268 & 1.471 \\
              & &           &     &    &      &      &      &         &         &       & 160 & 21.66 & ---    & ---   & ---   & ---     & 5.094 & 1.458 \\
 \hline \hline
\end{tabular}
\end{table*}

\begin{table*}[htp]
\caption{Same as Table~\ref{tab:H4_aligned} for spin-antialigned configurations.} 
\label{tab:anti-aligned}
\begin{tabular}{ccccccccccccccccccc}
    \toprule
EOS & $q$ & $M$ & $\kappa_2^A$ & $\kappa_2^B$ & $\chi^A$ & $\chi^B$ &  $\overline{\chi}$ & $M_{\rm ADM}$ & $J_{\rm ADM}$ & $M \omega^0_{2,2}$ & $n$ & $t_{\rm mrg}$ &  $t_{\rm coll}$ & $M_{\rm BH}$ & $\chi_{\rm BH}$ & $M_{\rm disk}$ & $e^{\rm mrg}_{\rm GW}$ & $L^{\rm mrg}_{\rm GW}$ \\
& & $[M_{\odot}]$ & & & & & & $[M_{\odot}]$ & $[M^2_{\odot}]$ & $[10^{-2}]$ & & $[\rm ms]$ &  $[\rm ms]$ & $[M{\odot}]$ & & $[10^{-2} M_{\odot}]$ & $[10^{-2} M_{\odot}]$ & $[10^{-3}]$ \\\hline \hline
SLy & 1.625  & 2.7 & 14  & 77  & -0.12 &  0.195 & 0.075  & 2.67998 & 6.77426 & 4.399 & 96  & 19.59 & ---    & ---   & ---   & ---     & 5.615 & 1.523 \\
    &        &     &     &     &       &        &        &         &         &       & 128 & 19.57 & ---    & ---   & ---   & ---     & 5.356 & 1.625 \\
    &        &     &     &     &       &        &        &         &         &       & 160 & 19.95 & ---    & ---   & ---   & ---     & 5.664 & 1.766 \\
    &        &     &     &     & 0.12  & -0.195 & -0.075 & 2.67985 & 7.18424 & 4.392 & 96  & 18.39 & ---    & ---   & ---   & ---     & 5.065 & 0.997 \\
    &        &     &     &     &       &        &        &         &         &       & 128 & 18.71 & ---    & ---   & ---   & ---     & 5.138 & 1.008 \\  
    &        &     &     &     &       &        &        &         &         &       & 160 & 18.55 & ---    & ---   & ---   & ---     & 4.928 & 0.966 \\
H4  & 1.625  & 2.9 & 23  & 135 & -0.12 & 0.195  & 0.075  & 2.87624 & 7.69938 & 4.860 & 96  & 12.99 & 1.36   & 2.65  & 0.62  & 7.94    & 4.898 & 0.884 \\
    &        &     &     &     &       &        &        &         &         &       & 128 & 12.85 & 1.66   & 2.65  & 0.63  & 7.69    & 4.929 & 0.902 \\
    &        &     &     &     &       &        &        &         &         &       & 160 & 13.27 & 1.31   & 2.66  & 0.63  & 7.22    & 4.947 & 0.913 \\
    &        &     &     &     & 0.12  & -0.195 & -0.075 & 2.87620 & 8.03535 & 4.857 & 96  & 10.26 & ---    & ---   & ---   & ---     & 4.436 & 0.523 \\
    &        &     &     &     &       &        &        &         &         &       & 128 & 10.48 & ---    & ---   & ---   & ---     & 4.468 & 0.541 \\
    &        &     &     &     &       &        &        &         &         &       & 160 & 10.58 & ---    & ---   & ---   & ---     & 4.478 & 0.537 \\
\hline \hline
\end{tabular}
\end{table*}

\end{document}